\documentclass[conference]{IEEEtran}

\ifCLASSINFOpdf
  \usepackage[pdftex]{graphicx}
  \graphicspath{{graphics/}}
  \DeclareGraphicsExtensions{.pdf}
\else
\fi
\usepackage[cmex10]{amsmath}

\begin{document}
%
\title{Detect \& Describe: \\
Deep learning of bank stress in the news}

\author{\IEEEauthorblockN{Samuel R\"onnqvist}
\IEEEauthorblockA{Turku Centre for Computer Science -- TUCS\\
Department of Information Technologies\\
\AA bo Akademi University, Turku, Finland\\
sronnqvi@abo.fi}
\and
\and
\IEEEauthorblockN{Peter Sarlin}
\IEEEauthorblockA{Department of Economics\\
Hanken School of Economics, Helsinki, Finland\\
RiskLab Finland\\
Arcada University of Applied Sciences\\
peter@risklab.fi}}

\maketitle

\begin{abstract}
News is a pertinent source of information on financial risks and stress
factors, which nevertheless is challenging to harness due to the sparse
and unstructured nature of natural text. We propose an approach based
on distributional semantics and deep learning with neural networks
to model and link text to a scarce set of bank distress events. Through
unsupervised training, we learn semantic vector representations
of news articles as predictors
of distress events. The predictive model that we learn can signal coinciding
stress with an aggregated index at bank or European level, while crucially
allowing for automatic extraction of text descriptions of the events,
based on passages with high stress levels. The method offers insight
that models based on other types of data cannot provide, while offering
a general means for interpreting this type of semantic-predictive
model. We model bank distress with data on 243 events and 6.6M news
articles for 101 large European banks.
\end{abstract}




%
\IEEEpeerreviewmaketitle

\section{Introduction}

The global financial crisis has triggered a large number of regulatory
innovations, yet little progress has occurred with respect to timely
information on bank vulnerability and risk. This paper provides a
text-based approach for identifying and describing bank distress using
news.

Prediction of bank distress has been a major topic both in the pre-
and post-crisis era. Many efforts are concerned with identifying the
build-up of risk at early stages, oftentimes relying upon aggregated
accounting data to measure imbalances (e.g., \cite{cole_predicting_1998,mannasoo_explaining_2009,betz2014predicting}).
Despite their rich information content, accounting data pose two major
challenges: low reporting frequency and long publication lags. A more
timely source of information is the use of market data to indicate
imbalances, stress and volatility (e.g., \cite{Groppetal2006,milne2014}).
Yet, market prices provide little or no descriptive information per
se, and only yield information about listed companies or companies'
traded instruments (such as Credit Default Swaps). This points to
the potential value of text as a source for understanding bank distress.

The literature on text-based computational methods for measuring risk
or distress is still scant. For instance, Nyman et al. \cite{Gregoryetal2014}
analyze sentiment trends in news narratives in terms of excitement/anxiety
and find increased consensus to reflect pre-crisis market exuberance,
while Soo \cite{soo2013quantifying} analyses the connection between
sentiment in news and the housing market. Both rely on manually-crafted
dictionaries of sentiment-bearing words. While such analysis can provide
interesting insight as pioneering work on processing expressions in
text to study risk, the approach is limiting as dictionaries are cumbersome
to adapt to specific tasks and generally incomplete.

Data-driven approaches, such as Wang \& Hua \cite{wang2014copula}
predicting volatility of company stocks from earning calls, may avoid
these issues. Their method, although allegedly providing good predictive
performance gains, offers only limited insight into the risk-related
language of the underlying text data. It also leaves room for further
improvements with regards to the semantic modeling of individual words
and sequences of words, which we address. Further, Lischinsky \cite{lischinsky2011discourse}
performs a crisis-related discourse analysis of corporate annual reports
using standard corpus-linguistic tools, including some data-driven
methods that enable exploration based on a few seed words. His analysis
focuses extensively on individual words and their qualitative interpretation
as part of a crisis discourse. Finally, R\"onnqvist \& Sarlin \cite{RonnqvistSarlin2015}
construct network models of bank interrelations based on co-occurrence
in news, and assess the information centrality of individual banks
with regards to the surrounding banking system.

We focus on a purely data-driven approach to \emph{detect} and \emph{describe} risk, 
in terms of a quantitative index and
extracted descriptions of relevant events. In particular, we demonstrate
this by learning to predict coinciding bank stress based on news,
where a central challenge is to link the sparse and unstructured text
to a small set of reference events. To this end, we demonstrate a
deep learning setup that learns semantic representations of text data
for a predictive model. We train the model to provide a coinciding
distress measure, while, most importantly, connecting text and distress
to provide descriptions. These text descriptions help explain the
quantitative response of the predictive model and allow insight into
the modeled phenomenon. The method is readily adaptable to any phenomenon
by selecting the type of reference events for training.

In the following section, we discuss the data we use to demonstrate
our approach to the study of stress. The deep learning setup, including
semantic modeling, predictive modeling, extraction of descriptions
and the related stress index is explained in Section 3. Finally, we
report on our experiments and reflect on the results in Section 4.

\section{Data}

The modeling in this paper is founded on connecting two types of data,
text and event data, by chronology and entities. The event data set
covers data on large European banks (entities), spanning periods before,
during and after the global financial crisis of 2007--2009. We include
101 banks from 2007Q3--2012Q2, for which we observe 243 distress events.
Following Betz et al. \cite{betz2014predicting}, the events include
government interventions and state aid, as well as direct failures
and distressed mergers. 

The text data consist of news articles from Reuters online archive
from the years 2007 to 2014 (Q3). The data set includes 6.6M articles
(3.4B words). As a first step towards linking bank distress events
and relevant news reporting, we identify mentions of the target banks.
Bank name occurrences are located using a set of patterns defined
as regular expressions that cover common spelling variations and abbreviations.
The patterns have been iteratively developed against the data to increase
accuracy, with the priority of avoiding false positives (in accordance
to \cite{RonnqvistSarlin2015}). Scanning the corpus, 262k articles
are found to mention any of the 101 target banks.

Each matching article is cross-referenced against the event data in
order to cast the article as distress-coinciding or not. An article
is considered distress-coinciding for a given bank if the bank is
mentioned and an event occurred within a month of the article's day
of publication. An article is considered not to coincide if it is
published at least three months from an event, whereas articles between
one and three months off are discarded to avoid ambiguity. The training
data is organized as tuples of bank name, article, publication date
and the assigned distress label. The publication dates are subsequently
aggregated monthly for the analysis of distress levels of banks over
time.

\section{The semantic deep learning setup}

Characterized in part by the deep, many-layered neural networks, a
prevailing idea of the deep learning paradigm is that machine learning
systems can become more accurate and flexible when we allow for abstract
representations of data to be successively learned, rather than handcrafted
through classical feature engineering. For a recent general survey
on deep learning confer Schmidhuber \cite{schmidhuber2015deep}, and
for a more explicit discussion of deep learning in natural language
processing Socher \& Manning \cite{manning2013deeplearning}.

While manually designed features help bring structure to the learning
task through the knowledge they encode, they often suffer problems
of being over-specified, incomplete and laborious to develop. Especially
regarding natural language processing, this limits the robustness
of text mining systems and their ability to generalize across domains,
tasks and languages. By exploiting statistical properties of the data,
features can be learned in an unsupervised fashion instead, which
allows for large-scale training not limited by the scarcity of annotated
data. Such intensively data-driven, deep learning approaches have
in recent years led to numerous breakthroughs in a range of application
domains from computer vision to natural language processing, where
a common theme is the use of unsupervised pre-training to effectively
support supervised learning of deep networks \cite{schmidhuber2015deep}.
We apply the same idea in modeling bank stress in news, as discussed
in the following.

\subsection{Modeling}

We are interested in modeling the semantics of words and complete
news articles to obtain suitable representations for predicting distress.
At the word level, distributional semantics exploits the linguistic
property that words of similar meaning tend to occur in similar contexts
\cite{harris1954distributional}. Modeling of word contexts yields
distributed representations of word semantics as vectors, which allow
measuring of semantic similarities and detecting analogies without
supervision, given substantial amounts of text \cite{Schutze:1992:DM:147877.148132,schutzePedersen1995irwordsenses,mikolov2013efficient}.
These word vectors provide a continuous semantic space embedding where
the symbolic input of words can be compared quantitatively,
thus supporting both the prediction task of this paper and a multitude
of other natural language processing tasks (e.g., tagging, parsing
and relation extraction \cite{manning2013deeplearning}). 

While traditionally modeled by counting of context words, predictive
models have eventually taken the clear lead in terms of performance
\cite{baroni2014don}. Neural network language models in particular
have proved useful for semantic modeling, and are especially practical
to incorporate into deep learning setups due to their dense vectors
and the unified neural framework for learning. Mikolov et al. \cite{mikolov2013efficient}
have put forward an efficient neural method that can learn highly
accurate word vectors as it can train on massive data sets in practical
time (a billion words in the order of a day on standard architecture).
Subsequently, Le \& Mikolov \cite{le2014distributed} extended the
model in order to represent compositional semantics (cf. \cite{mitchell2010composition})
of sequences of words, from sentences to the length of documents,
which they demonstrated to provide state-of-the-art performance on
sentiment analysis of movie reviews. Analogous to that task, we employ
their distributed memory method to learn document vectors for news
articles and use them for learning to predict coinciding distress,
as a type of risk sentiment analysis guided by the event data we provide.

\begin{figure}
\begin{center}
\includegraphics[width=0.9\columnwidth]{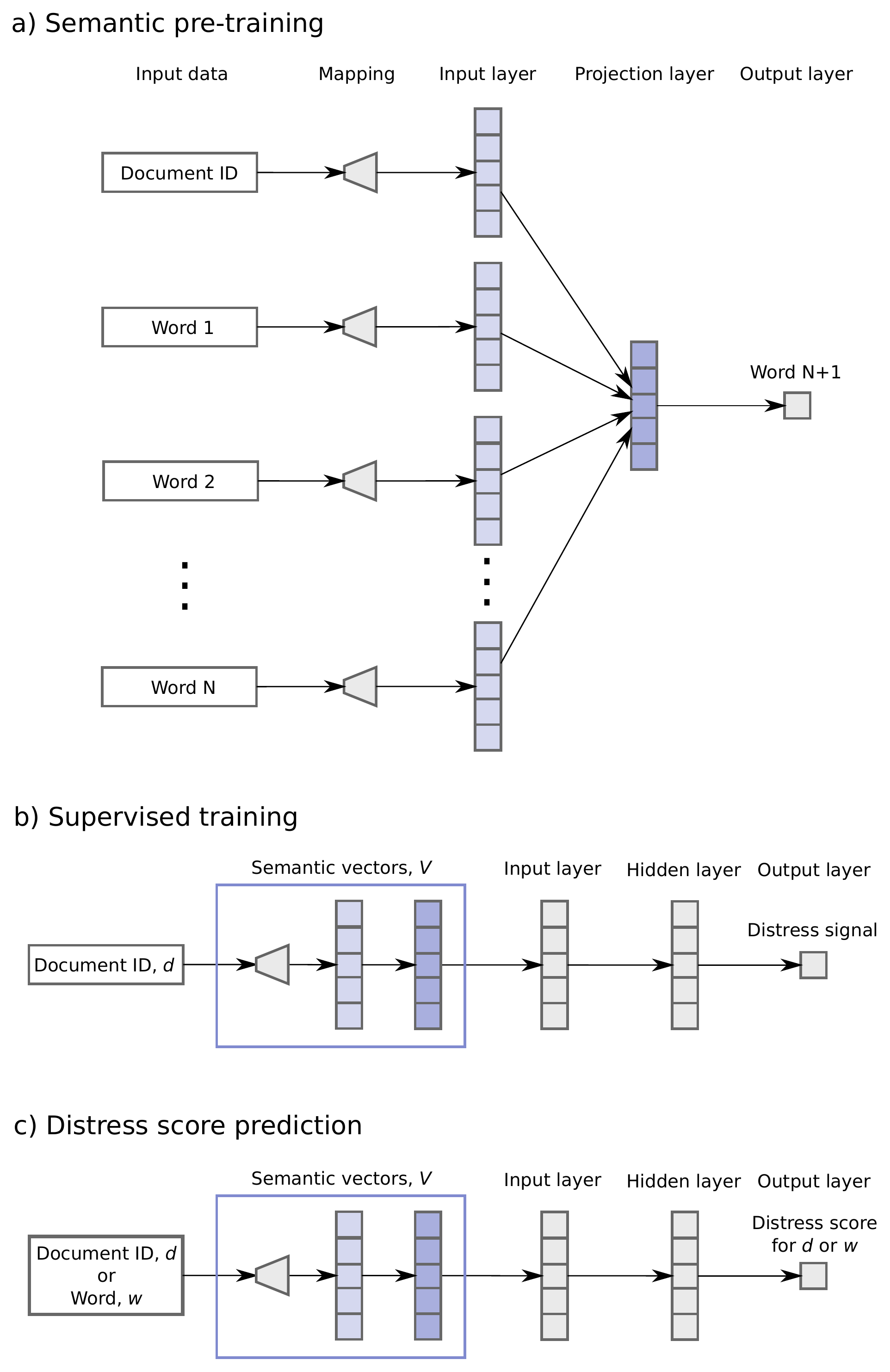}
\end{center}
\protect\caption{Deep neural network setup for (a) pre-training of semantic vectors,
(b) supervised training against distress event signal $e$, and (c) prediction
of distress score $M(V)$ for a document or word.}
\label{fig_deep_setup}
\end{figure}

Our deep neural network for predicting distress from text, outlined
in Fig.~\ref{fig_deep_setup}, is trained in two steps: through learning
of document vectors as pre-training (\ref{fig_deep_setup}a), followed
by supervised learning against the distress signal (\ref{fig_deep_setup}b).
The use of the distributed memory model in the first step is explained
in the following.

The modeling of word-level semantics works by taking a sequence of
words as input and learning to predict the next word (e.g., the 8\textsuperscript{th}
in a sequence), using a feed-forward topology where a projection layer
in the middle provides the semantic vector once its weights have been
learned. The projection layer provides a linear combination that enables
efficient training on large data sets, which is important in achieving
accurate semantic vectors. In addition, document vectors include the
document ID as input, functioning as a memory for the model that allows
the vector to capture the semantics of continuous sequences rather
than only single words; the document ID in fact can be though of as
an extra word representing the document and informing the prediction
of the next word. Formally, the pre-training step seeks to maximize
the average log probability:

\[
\frac{1}{T-N}\sum_{t=N+1}^{T}\mathrm{log}\, p(w_{t}|d,w_{t-N},...,w_{t-1})
\]
over the sequence of training words $w_{1},w_{2},...,w_{T}$ in document
$d$ with word context size $N$. In the neural network, an efficient
binary Huffman coding is used to map document IDs and words to the
input layer, which imposes a basic organization of words by frequency. 

The second modeling step (Fig. \ref{fig_deep_setup}b) is a normal
feed-forward network fed by the document vectors $V_{d}$ (pertaining
to the set of documents $D$), which we train by stochastic gradient
descent and backpropagation \cite{rumelhart1986learning} to predict
distress events $e\in\left\{ 0,1\right\}$. Hence, the objective is to maximize the average
log probability:

\[
\frac{1}{|D|}\sum_{d\in D}\mathrm{log}\, p(e_{d}|V_{d})
\]

Compared to sequential text, the document vector is a practical fixed-size
representation suitable as input to a feed-forward network. For each
word, the input text originally has a dimensionality equal to the
vocabulary size (typically millions of words), but the semantic modeling
provides reduction to the size of the vector (typically 50--1000).
Both these aspects help train the model against a signal
corresponding to a comparatively tiny number of events. In our experiments,
document vectors are learned based on all bank-related articles and
the predictive network is trained on individual pairs of document
vector and binary distress signal, matched together based on date
and bank name occurrence as discussed in Section 2.

\subsection{Stress index and extraction of its description}

As the network in step two has been trained and the hyperparameters
optimized by validation, it can be applied to articles and the posterior
probability used as a stress score. The scores are aggregated over
articles per bank and period by their mean to provide a stress index
$I:p\times b\rightarrow[0,1]$:

\begin{equation}
I(p,b)=\frac{1}{|D_{p,b}|}\sum_{d\in D_{p,b}}M(V_{d})
\end{equation}
over the documents $D_{p,b}$ that mention bank $b$ in period $p$,
where $M(V)=p(e=1|V)$ gives the posterior probability of the trained neural network model.

Guided by the index, we can choose banks and periods for closer inspection,
where the text data and our semantic-predictive model play the important
role of providing descriptions of events that the index reflects.
We use the trained semantic representations and their predicted signal
strength to find sections in the articles that are strongly linked
to stress. Fig. \ref{fig_deep_setup}c illustrates how the trained
network is used to obtain both article-level stress scores $M(V_{d})$ and word-level
stress scores $M(V_{w})$, which are combined for the extraction of descriptions.
The extraction operates based on a weighting of words defined as:

\begin{equation}
x_{d,w}=M(V_{d})\cdot M(V_{w})\cdot f_{d,w}
\end{equation}
where $V$ maps document\emph{ }$d$ as well as word\emph{ }$w$\emph{
}to their respective vectors\emph{. }The word count in a given document
is defined by $f$. 

The intuition is that a word is descriptive of distress if: (1) it
occurs in a document that as a whole produces a high stress score,
(2) the individual word has a vector that in itself produces a high
score, and (3) the word is prevalent in the discussion of the article.
While individual words are atomic and interpretation-friendly, the
document-level score incorporates a useful contextualization with
respect to distress. As word vectors and document vectors reside in
the same space and are directly comparable, word vectors can be fed
to the predictive model, even though it is trained only on document
vectors. The model operating meaningfully on word vectors as well
as document vectors is a most useful side effect. By comparison, searching
for words with vectors similar to high-scoring document vectors yields
significantly noisier results, thus exploiting the word vectors directly
through stress prediction offers superior guidance in description
extraction. Moreover, word frequency improves the results by accounting
for word repetition in articles and stop word filtering discards obviously
uninformative keywords.

We represent descriptions in two ways: as individual keywords and excerpts of
text. Keywords are extracted directly by ranking according to word
score $x$, and may be semantically organized using word-vector-based
similarity for improved readability. Yet, single words may be difficult
to interpret without context, which motivates the extraction of complete
excerpts from the articles. A fixed-length sliding window over words
collects excerpt candidates, which are ranked by their total word
score and filtered by word overlap to avoid redundant excerpts down
the rank. Optionally, the occurrence of the target bank name may be
required or up-weighted in a excerpt in order to focus on the discussion
most closely related to the bank of interest.

The excerpt candidates and weighted words can be aggregated in different
ways before ranking and filtering, to support various types of analysis.
Aggregation by period highlights the most prominent distress-related
events of each cross section (see Section 4.2), while aggregation
by period and bank provides more focused descriptions relating to the
distress discourse of specific entities over time (see Section 4.3).

The weighting in Eq. 2 encodes the relevance of words with regards
to bank distress as the target phenomenon. In the experiments discussed
next, we demonstrate the utility of the excerpts in providing insight
into events surrounding particular banks over time, as the excerpts
and keywords in concord highlight and describe driving forces behind
the stress index.

\section{Experiments}

As input for the first step of training the document vectors, all
articles that mention any of the target banks are used. Corresponding
document vectors are learned for each of the 262k articles, trained
over sequences of their in total 210M words, that capture the semantics
specifically of reporting related to our banks of interest. The train
set could be expanded to include other text, too, as more training
data (often up to several billion words) generally provide semantic
models with higher accuracy and broader coverage. However, it is not
clear that a broader coverage in training data would benefit the current
task, while it would substantially increase the computational time
and space complexities. In practice, the memory required to train
on large numbers of documents, e.g., exceeding a million, is likely
to be a bottleneck on standard PC hardware. We train a vector of length
400 using a context of 8 words (similar to the sentiment experiment
of \cite{le2014distributed}).

\subsection{Predictive modeling and evaluation}

Following the semantic pre-training, we train a predictive neural
network model with 3 layers. The input layer has 400 nodes, corresponding
to the semantic vectors and a single output corresponding to distress.
As described in Section 2, a set of tuples are complied as data to
learn a predictive model of distress based on articles. The set consists
of 211k cases, 10.7\% of which are labeled as distress-coinciding
following our matching procedure. The skewed classes require care
in evaluation, as does the imbalanced preference between types of
errors: we concider missing an event much worse than incorrectly signaling
one.

We evaluate the performance of the predictive model to asses the quality
of the stress index it will produce, and importantly to provide in
extension a quality assurance for the information content of the descriptions
we extract. We use the relative Usefulness measure ($U_{r}$) by Sarlin
\cite{Sarlin2013b}, as it is commonly used in distress prediction
and intuitively incorporates both error type preference ($\mu$) and
relative performance gain of the model. Based on the combination of
negative/positive observations ($\mathrm{obs}\in\left\{0,1\right\}$) and negative/positive
predictions ($\mathrm{pred}\in\left\{0,1\right\}$), we obtain the cases of true
negative ($TN\equiv obs=0\wedge pred=0$), false negative ($FN\equiv obs=1\wedge pred=0$),
false positive ($FP\equiv obs=0\wedge pred=1$) and true positive
($TP\equiv obs=1\wedge pred=1$), for which we can estimate probabilities
when evaluating our predictive model. Further, we define the baseline
loss $L_{b}$ to be the best guess according to prior probabilities
$p(\mathrm{obs})$ and error preferences $\mu$ (Eq. 3) and the model
loss $L_{m}$ (Eq. 4):

\begin{equation}
L_{b}=\mbox{min}\begin{cases}
\mu\cdot p(\mathrm{obs}=1)\\
(1-\mu)\cdot p(\mathrm{obs}=0)
\end{cases}
\end{equation}

\begin{equation}
L_{m}=\mu\cdot p(FN)+(1-\mu)\cdot p(FP)
\end{equation}

From the loss functions we derive Usefulness in absolute ($U_{a}$) and relative terms ($U_{r}$):

\begin{equation}
U_{r}=\frac{U_{a}}{L_{b}}=\frac{L_{b}-L_{m}}{L_{b}}
\end{equation}

While absolute Usefulness $U_{a}$ measures the gain vis-\`a-vis
the baseline case, relative Usefulness $U_{r}$ relates gain to
that of a perfect model (i.e., Eq. 5 with $L_{m}=0\Rightarrow U_{a}=L_{b}$).
Usefulness functions both as a proxy for benchmarking the model (testing)
and to optimize its hyperparameters (validation). For reference, Table
\ref{eval_table} also reports the performance as the in text mining
widely used $F$-score\cite{vanRijsbergen1979} (based on $\mathrm{precision}=p(\mathrm{obs}=1|\mathrm{pred}=1)$
and $\mathrm{recall}=p(\mathrm{pred}=1|\mathrm{obs}=1)$):

\begin{equation}
F_{\beta}=(1+\beta^{2})\cdot\frac{\mbox{precision\ensuremath{\cdot}recall}}{(\beta^{2}\cdot\mbox{precision})+\mbox{recall}}
\end{equation}
which similar to Usefulness can account for varying preferences by
its $\beta$ parameter, although not gain. The $F_{\beta}$-score
assigns $\beta$ times as much importance to recall as to precision
(i.e., preference for completeness over exactness)\cite{vanRijsbergen1979},
which is analogous to but not directly transferable to the $\mu$
parameter in the Usefulness measure. While the $F$-score is commonly
seen to maximize completeness versus exactness of true positives,
the parameter can also be seen as a priority to minimize false negatives
versus false positives (FN prioritized over FP when $\beta>1$). As
a heuristic, we map the balanced, standard $F_{1}$-score with $\beta=1$
to $U_{r}$ with $\mu=0.5$, and match deviations from these preferences
according to $\beta=\mu/(1-\mu)$.

\begin{table}
\begin{tabular}{ccccccccc}
$\mu$ & $\bar{U}_{r}(\mu)$ & $\sigma_{U}$ & $\bar{F}_{\beta}$ & $\sigma_{F}$ & $\bar{T\! N}$ & $\bar{F\! N}$ & $\bar{F\! P}$ & $\bar{T\! P}$\tabularnewline
\hline 
\hline 
0.1 & -33.34 & 2.47 & 0.070 & 0.03 & 18823 & 2249 & 0 & 2\tabularnewline
0.2 & -14.41 & 1.08 & 0.016 & 0.01 & 18823 & 2249 & 1 & 2\tabularnewline
0.3 & -8.094 & 0.62 & 0.023 & 0.01 & 18819 & 2242 & 4 & 8\tabularnewline
0.4 & -4.937 & 0.39 & 0.036 & 0.01 & 18807 & 2225 & 16 & 26\tabularnewline
0.5 & -3.044 & 0.25 & 0.037 & 0.02 & 18787 & 2208 & 36 & 43\tabularnewline
0.6 & -1.781 & 0.16 & 0.066 & 0.02 & 18694 & 2142 & 130 & 109\tabularnewline
0.7 & -0.879 & 0.09 & 0.114 & 0.02 & 18455 & 2023 & 368 & 228\tabularnewline
0.8 & -0.203 & 0.04 & 0.229 & 0.03 & 17503 & 1740 & 1321 & 511\tabularnewline
0.85 & 0.147 & 0.02 & 0.397 & 0.05 & 15462 & 1331 & 3362 & 919\tabularnewline
0.9 & \textbf{0.271} & 0.01 & \textbf{0.713} & 0.03 & 10306 & 577 & 8517 & 1673\tabularnewline
0.95 & 0.146 & 0.01 & 0.934 & 0.01 & 4873 & 112 & 13950 & 2139\tabularnewline
\end{tabular}
\\
\protect\caption{Cross-validated predictive performance as relative Usefulness and $F$-score over
preferences between types of error ($\mu$) and recall/precision ($\beta$).}
\label{eval_table}
\end{table}

Based on relative Usefulness, we find the optimal network (20 hidden
nodes) and hyperparameters for the stochastic gradient decent algorithm
to train its weights. For evaluation, we trained the network by randomized
10-fold cross validation with one fold for validation and one for
testing. Table \ref{eval_table} reports the performance of the optimal
models on the test set. The evaluation yielded an area under the ROC
curve of 0.710 with a standard deviation $\sigma=0.006$. Following
previous studies \cite{betz2014predicting,Peltonenetal2015}, we make
use of a skewed preference $\mu=0.9$ (i.e., missing a crisis is 9
times worse than falsely signaling one). From the viewpoint of policy,
highly skewed preferences are particularly motivated when a signal
leads to an internal investigation, and reputation loss or other political
effects of false alarms need not be accounted for. We conclude that
at $\mu=0.9$ the model has decent predictive performance by capturing
27\% of available Usefulness (cf. \cite{betz2014predicting,Peltonenetal2015}).
While the model is not robust to low levels of $\mu$, we can see
in Table \ref{eval_table} that Usefulness is positive for $\mu$
around 0.9.

\begin{figure*}
\includegraphics[width=0.99\textwidth]{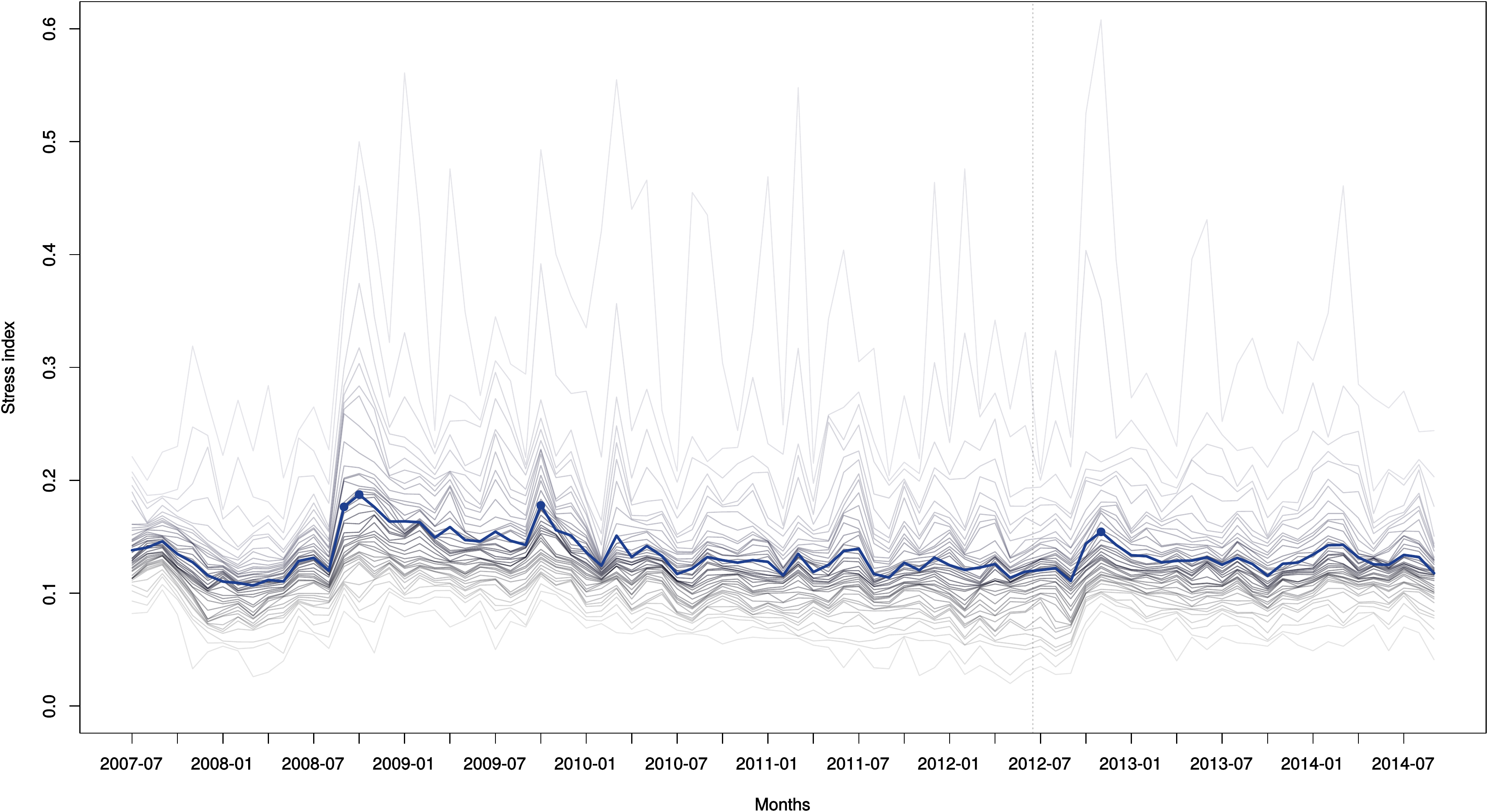}

\protect\caption{Distress index distribution over time for all banks (blue line indicates
mean, faded lines represent every 2.5 percentiles). Periods from July 2012
onward (right of dashed line) are outside the event sample. Key periods
of interest are marked by points.}

\label{fig_index}
\end{figure*}

\subsection{Stress index, descriptions and interpretation}

Having trained the network and evaluated its predictive performance,
we can reliably extract stress indices for banks over time, as discussed
in Section 3.2. In order to provide an overview of the indices of
all 101 banks, Fig. \ref{fig_index} shows the dynamics of the mean
index and percentiles for the monthly cross sections. The time span
July 2007 to June 2012 is covered by the event data (where a random
80\% of cases were used for training, 10\% for validation, and 10\%
for testing), and the span from July 2012 onward is completely out-of-sample.
The indices show a sharp increase in September 2008 and further rises in October, during
the outbreak of the financial crisis, followed by a long tail of relatively
high stress, driven in part by a minority of the banks having high
values. Yet, no part of the cross section remains completely unaffected,
a pattern that is likewise pronounced in the sudden peak of October
2009. Similarly, a noteable jump in the entire cross section occurs
in October 2012. We describe these peaks by extracing top ranking
distress-related keywords and excerpts. The following discussion has
its basis in the excerpts in Tables II-V, which we select based on 
their relevance and diversity of specific topics, generally among the top-20 or top-40 ranked.

\begin{table}
\begin{tabular}{p{0.2cm}p{8cm}}

1) & \emph{fortis, rescue, markets, european, belgian-dutch, crisis, sunday,
nationalisation}\\
 & ``...crisis has kept markets on tenterhooks by forcing European authorities
to rescue troubled banks. Belgian-Dutch group Fortis FOR.BR underwent
nationalisation on Sunday after emergency...''\footnotemark
\tabularnewline
2) & \emph{european, rbs, banking, stocks, banks, ubsn, bank, nationalisation}\\
& ``...part nationalisation of two major European banks battered banking
stocks, with Royal Bank of Scotland (RBS.L) falling 16.8 percent,
Swiss bank UBS (UBSN.VX) losing 13.6...''\footnotemark
\tabularnewline

3) & \emph{markets, stock, bailout, crisis, investors, conditions, potential,
ease}\\
& ``...would ease the financial crisis. The proposed government bailout
has become a stark reminder to investors and potential homebuyers
of worsening economic conditions. Stock markets...''\footnotemark
\tabularnewline

4) & \emph{britain, lender, hbos, deal, week, government, ease, competition}\\
& ``...week, Lloyds rescued Britain's biggest mortgage lender HBOS
in a \$22 billion takeover as the government swept aside competition
rules to ease the deal through.''\footnotemark
\tabularnewline

5) & \emph{stock, market, britain, short, stocks, toxic, bank, debt}\\
& ``...toxic mortgage-related debt and Britain cracked down on short
selling of bank stocks. The impact was immediate and dramatic, driving
the U.S. stock market...''\footnotemark
\tabularnewline

6) & \emph{hypo, monday, guarantees, hrxg, lender, german, spokesman, banks}\\
& ``(\$51.21 billion) in credit guarantees to cash-strapped German
lender Hypo Real Estate HRXG.DE, a Finance Ministry spokesman said
on Monday. 'No foreign banks took part...'''\footnotemark
\end{tabular}
\\
\protect\caption{Selected top-ranked keywords and excerpts for September 2008}
\end{table}

\addtocounter{footnote}{-6} 
\stepcounter{footnote}\footnotetext{http://www.reuters.com/article/2008/09/29/markets-europe-stocks-open-idUSLT42646320080929}
\stepcounter{footnote}\footnotetext{Ibid. /2008/09/29/markets-europe-stocks-closer-idUKLT49969620080929}
\stepcounter{footnote}\footnotetext{Ibid. /2008/09/24/usa-economy-mortgages-idUSN2444124620080924} 
\stepcounter{footnote}\footnotetext{Ibid. /2008/09/22/ukbanks-research-jpmorgan-idUSBNG9691820080922} 
\stepcounter{footnote}\footnotetext{Ibid. /2008/09/19/financial-idUSHKG9362820080919} 
\stepcounter{footnote}\footnotetext{Ibid. /2008/09/29/hyporeal-credit-foreigners-idINBAT00237520080929}

\begin{table}
\begin{tabular}{p{0.2cm}p{8cm}}

1) & \emph{fortis, dutch, european, million, abn, euros, sell, week}\\
& ``...euros' (\$970 million) worth of Dutch ABN AMRO assets to Deutsche
Bank last week. Fortis needed to sell some operations to meet European
Commission antitrust...''\footnotemark
\tabularnewline

2) & \emph{dutch, cash, market, euro, sunday, losses, injection, ing}\\
&``...banks booked losses on the U.S. housing market. On Sunday, the
Dutch government agreed a 10 billion euro cash injection to ING to
shore...''\footnotemark
\tabularnewline

3) & \emph{fortis, dutch, european, crisis, euro, group, zone, government}\\
&``...European Central Bank and euro zone finance ministers to discuss
a response to the global financial crisis. 'Fortis Group has sold
to the Dutch government...'''\footnotemark
\tabularnewline

4) & \emph{crisis, cash, help, government, take, barclays, bank, royal}\\
&``...take government cash. Rivals Royal Bank of Scotland, Lloyds
and HBOS are all taking billions in taxpayers' funds to help weather
the financial crisis. Barclays...''\footnotemark
\tabularnewline

5) & \emph{markets, new, coordinated, cuts, banks, bank, investors, central}\\
&``...investors feared that Wednesday's coordinated interest rate
cuts by global central banks will not untangle credit markets and
avert recession. The Bank of New York...''\footnotemark
\tabularnewline

6) & \emph{rescue, european, institutions, banks, below, details, euros,
plans}\\
&``...combined pledges of capital injections into European banks and
financial institutions to over 1 trillion euros. Below are details
of the financial rescue plans already...''\footnotemark
\tabularnewline

7) & \emph{markets, sept, stock, take, bailout, toxic, off, restore}\\
&``...take toxic mortgage assets off the books of financial companies
to restore financial stability. News of the bailout plan helps world
stock markets soar. Sept. ...''\footnotemark
\tabularnewline
\end{tabular}
\\
\protect\caption{Selected top-ranked keywords and excerpts for October 2008}
\end{table}

\addtocounter{footnote}{-7} 
\stepcounter{footnote}\footnotetext{Ibid. /2008/10/10/sppage012-la190235-oisbn-idUSLA19023520081010}
\stepcounter{footnote}\footnotetext{Ibid. /2008/10/20/us-socgen-shares-idUSTRE49J1WG20081020}
\stepcounter{footnote}\footnotetext{Ibid. /2008/10/06/sppage012-l3713090-oisbn-idUSL371309020081006}
\stepcounter{footnote}\footnotetext{Ibid. /2008/10/31/britain-barclays-idUSLV68208720081031}
\stepcounter{footnote}\footnotetext{Ibid. /2008/10/08/markets-stocks-adrs-idUSN0839418020081008}
\stepcounter{footnote}\footnotetext{Ibid. /2008/10/14/us-financial-rescues-factbox-\\idUSTRE49D61Y20081014}
\stepcounter{footnote}\footnotetext{Ibid. /2008/10/02/financial-idUSN0153441520081002}

\begin{table}
\begin{tabular}{p{0.2cm}p{8cm}}

1) & \emph{european, shares, losses, results, reuters, banking, stocks,
bank}\\
&``(Reuters) - European shares extended losses on Wednesday, as banking
stocks weighed following a surprise announcement of key quarterly
results by Deutsche Bank (DBKGn.DE).''\footnotemark
\tabularnewline
2) & \emph{dutch, ing, european, kbc, shares, reuters, bank, investors}\\
&``(Reuters) - Investors are shunning European bank shares after an
EU-imposed break-up and retrenchment of Dutch ING Groep (ING.AS) sparked
fears Belgium's KBC and UK...''\footnotemark
\tabularnewline
3) & \emph{dutch, aid, monday, stock, ing, repay, value, split}\\
&``...stock lost 18 percent of its value. ING said on Monday it would
split into two units, repay some of its Dutch state aid early...''\footnotemark
\tabularnewline
4) & \emph{fortis, abn, dutch, sell, nationalised, local, government, banking}\\
&``...sell some ABN AMRO assets in the Dutch small and medium enterprise
banking sector to address competition concerns. When the government
nationalised Fortis's local operations...''\footnotemark
\tabularnewline

\end{tabular}
\\
\protect\caption{Selected top-ranked keywords and excerpts for October 2009}
\end{table}

\addtocounter{footnote}{-4} 
\stepcounter{footnote}\footnotetext{Ibid. /2009/10/21/markets-europe-stocks-negative-\\idUSLL6262520091021}
\stepcounter{footnote}\footnotetext{Ibid. /2009/10/27/banks-commission-shares-idUKLC72878420091027}
\stepcounter{footnote}\footnotetext{Ibid. /2009/10/27/us-ing-idUSTRE59P0U020091027}
\stepcounter{footnote}\footnotetext{Ibid. /2009/10/19/abnamro-idUSLJ72227820091019}

\begin{table}
\begin{tabular}{p{0.2cm}p{8cm}}

1) & \emph{income, euro, results, zone, september, crisis, cut, benefit}\\
&``...euro zone crisis has pushed banks to cut back even more, even
as third-quarter results look set to benefit from better trading income
in September.''\footnotemark
\tabularnewline
2) & \emph{information, cuts, jobs, plans, suisse, swiss, ubs, technology}\\
&``...Swiss bank Credit Suisse may announce 1,000-2,000 cuts, Der
Sonntag newspaper reported. UBS plans to shed 900 jobs in information
technology,...''\footnotemark
\tabularnewline
3) & \emph{income, reuters, business, plans, tuesday, swiss, ubs, bank}\\
&``ZURICH (Reuters) - Swiss bank UBS unveiled plans on Tuesday to
fire 10,000 staff and wind down its fixed income business, returning
to its...''\footnotemark
\tabularnewline
4) & \emph{reuters, jobs, losses, likely, swiss, investment, banking, wednesday}\\
&``...investment banking jobs as early as Wednesday, a source familiar
with the situation told Reuters, with more job losses at the Swiss
bank likely to...''\footnotemark
\tabularnewline
5) & \emph{euro, european, bailout, high, buying, central, spain, support}\\
&``...high. Expectations that Spain will apply for a bailout, prompting
the European Central Bank to start buying its bonds, have helped support
the euro in...''\footnotemark
\tabularnewline
6) & \emph{euro, zone, reuters, bailout, losses, session, request, spain}\\
&``...previous session's losses, lifted by hopes that struggling Spain
will request a bailout which would lower its borrowing costs. Euro
zone sources told Reuters over...''\footnotemark
\tabularnewline
7) & \emph{euro, zone, likely, crisis, debt, recent, central, research}\\
&``'...the sovereign debt crisis,' Commerzbank economist Christoph
Weil wrote in a recent research note. Euro zone and UK central bankers
will likely leave policy unchanged...''\footnotemark
\tabularnewline
8) & \emph{euro, zone, european, single, week, request, investors, banking}\\
&``...week but uncertainty about when such a request might come has
made investors wary of driving the euro zone common currency much
higher. European leaders moved closer to establishing a single euro
zone banking...''\footnotemark
\tabularnewline

\end{tabular}
\\
\protect\caption{Selected top-ranked keywords and excerpts for October 2012}
\end{table}

\addtocounter{footnote}{-8} 
\stepcounter{footnote}\footnotetext{Ibid. /2012/10/24/us-ubs-jobs-idUSBRE89N0RL20121024}
\stepcounter{footnote}\footnotetext{Ibid. /2012/10/21/banks-switzerland-jobs-idUKL5E8LL10T20121021}
\stepcounter{footnote}\footnotetext{Ibid. /2012/10/30/us-ubs-restructure-idUSBRE89S0DM20121030}
\stepcounter{footnote}\footnotetext{Ibid. /2012/10/24/us-ubs-jobs-idUSBRE89N09620121024}
\stepcounter{footnote}\footnotetext{Ibid. /2012/10/22/markets-forex-idUSL1E8LMLMC20121022}
\stepcounter{footnote}\footnotetext{Ibid. /2012/10/15/markets-europe-stocks-idUSL5E8LF40K20121015}
\stepcounter{footnote}\footnotetext{Ibid. /2012/10/01/us-eurozone-unemployment-\\idUSBRE8900JQ20121001}
\stepcounter{footnote}\footnotetext{Ibid. /2012/10/19/markets-forex-idUSL1E8LJBZI20121019}

At a general level, the peak in \textbf{September 2008} can be seen
to relate to the overall distress in financial markets due to the
collapse of Lehman Brothers in mid-September. More specifically, the
discussion related to European banks can be seen to relate to the
nationalization of the Dutch and Luxembourgian arms of Fortis Bank.
The second excerpt indicates the impact it had on Europe-wide bankings
stocks, with Royal Bank of Scotland falling 16.8\%, UBS 13.6\% and
UniCredit 10.2\%, in addition to general indications of worsening
conditions. Likewise, the fourth excerpt highlights the impact on
UK mortgage lender HBOS and the fifth the general spillover between
the US and UK. The final excerpt highlights spread of distress to
the German lender Hypo Real Estate in late September.
The increase in the distress index in \textbf{October 2008 }confirms
the further spread of distress in Europe, as well as a continued discussion
around the early cases that were still to be resolved, such as the
Belgian subsidiary of Fortis that was sold to BNP Paribas in October.

The increase in the index in \textbf{October 2009} is again a general
peak in distress among European banks, as is pointed out by more widespread
losses mentioned in the first excerpt. This also coincides with the Greek legislative
elections, whereafter the de facto budget deficit is revealed to be
much larger than expected. Yet, as the second and third excerpts point
out, more specific distress relates to the EU-imposed break-up and
retrenchment of the Dutch ING Group, which set a precedent for other
bailed-out banks, such as Lloyds Banking Group, Royal Bank of Scotland
and KBC. In the same vein, the fourth excerpt mentioning Fortis points
to a still ongoing process of merging the Dutch insurance and banking
subsidiaries into ABN AMRO.

Finally, a noteworthy peak in the distress index occurs in \textbf{October
2012}. The first four excerpts all relate to cuts in the banking sector,
which highlights the impact of the crisis on the banking sector. Further,
excerpts five and six also mention risks with Spanish sovereign debt
and the potential need for a bailout. While the keywords related to
excerpt seven highlight debt sustainability and the impact of policy
changes, the eighth excerpt combines the topics of a Spanish bailout
and a Europe-wide banking supervisor. To this end, the excerpts for
October 2012 also bring together bank distress and debt sustainability
issues, thus highlighting the European bank-sovereign nexus.

The values at the top of the distribution appear rather unstable from
month to month, which reflects that different banks are being mentioned
over time and usually not persistently across months in distress contexts. By observing
increases and peaks in the index of an individual bank, we can locate
events of possible relevance to distress. The ability to
extract descriptions for these events then becomes useful in order
to discern what has happened in relation to the bank and distress.
To illustrate our approach for an individual bank, we turn in the
following section to a case study of Fortis Bank.

\begin{figure*}
\includegraphics[width=1\textwidth]{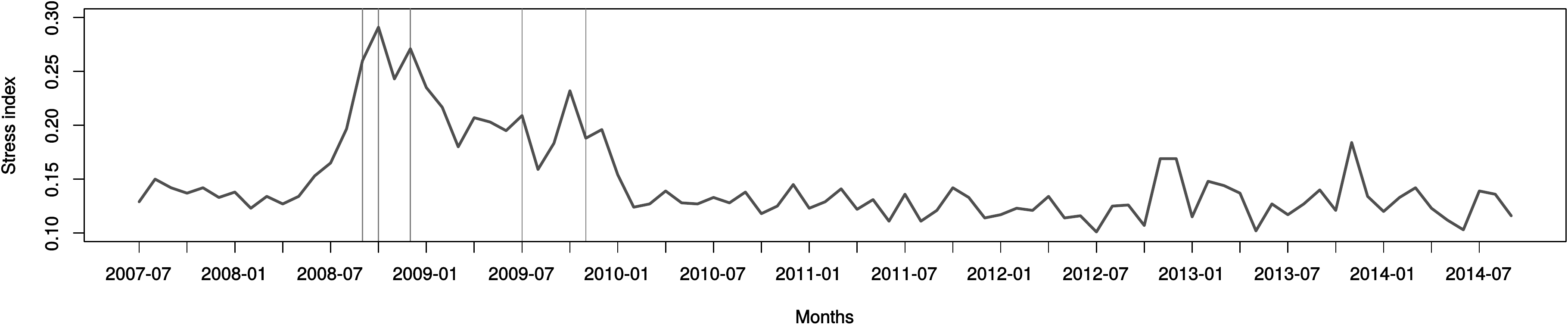}

\protect\caption{The distress index for Fortis from July 2007 to September 2014. Vertical lines
represent the start of a distress event.}

\label{fig_fortis}
\end{figure*}

\subsection{The case of Fortis Bank}

One of the early failures among European financial institutions occurred
to the Benelux-based Fortis. As was multiple times highlighted in
the above described top distress excerpts, Fortis and the rescue procedure
was at the core of the discussion in the crisis. This section focuses
on the evolution of the distress index for Fortis, as is shown in
Fig. \ref{fig_fortis}, and relates excerpts from Fortis-related
discussion to the peaks of their distress values and the realized
distress events (as shown with vertical lines in the figure). To start
with, we can observe that elevated values for the distress index coincide
with distress events. Further, a qualitative analysis of the stress
levels shows also that the index increases a few months prior to stress,
which indicates slight early-warning properties of the measure.

By assessing excerpts for all points in time when the distress index
took high values (breached 0.16), we provide a description from the
perspective of one individual bank. Fortis signaled distress already
in \textbf{July 2008} with a value of 0.17. The top ranked excerpts
relate to a range of different issues, without pointing out specific
causes of distress in Fortis. We interpret this as increases in general
concerns related to Fortis, but with no unanimity on the specific
problems at hand. 
The surge in \textbf{September 2008} is more specific
in nature. After mentions of Fortis potentially acquiring the insurer
Delta Lloyd ABN AMRO Verzekeringen Holding BV (excerpt 1), the second
excerpt notes that Fortis itself was forced to be nationalized by
the Benelux governments in late September. 

In \textbf{October 2008}, the excerpts relate to a divestment of Fortis'
Dutch assets in that the Dutch government purchased the banking and
insurance divisions in the Netherlands (excerpt 3), as well as rumours
of Fortis selling its Dutch banking arm to Deutsche Bank (excerpt
4). Despite the financial turmoil, a top excerpt (5) shows that the
Dutch unit of ABN AMRO that was earlier bought by Fortis showed an
increase in profits.  The distress index values continue to be high
until early 2010. As most of the discussion relates to similar issues
as at the early stages of the crisis, we only pinpoint a few interesting
instances. 
In \textbf{October 2009}, after a number of state interventions, a
top excerpt highlights that Deutsche Bank has finally agreed to buy
ABN AMRO assets from the Dutch government, which again should enable
a merger of the nationalized ABN and Fortis Bank Nederland. 

Interestingly, while a large number of excerpts still mention various
aspects of state interventions, a number of excerpts in \textbf{December
2009} highlight repayment of aid, such as BNP Paribas Fortis repaying
a total of €48.8 million to Belgium since the guarantees were put
in place in May 2009. Likewise, another top-ranked excerpt highlights
that BNP Paribas increases its forecast on savings from the integration
of former Fortis assets and a successful acquisition of the distressed
entity. While being handpicked excerpts of the top ranked excerpts,
this provides an idea of how large increases in the built stress index
can be accompanied by descriptive excerpts and eventually also the
original sources themselves, to help discern the relevant developments in play.

\begin{table}
\begin{tabular}{p{0.2cm}p{8cm}}

1) & September 2008: ``Belgian-Dutch financial group Fortis FOR.BRFOR.AS
buys Delta Lloyd ABN Amro Verzekeringen Holding BV, a Dutch insurance
company (notified Sept. 15/deadline Oct. 20...''\footnotemark
\tabularnewline
2) & September 2008: ``...crisis has kept markets on tenterhooks by forcing
European authorities to rescue troubled banks. Belgian-Dutch group
Fortis FOR.BR underwent nationalisation on Sunday after emergency...''\footnotemark
\tabularnewline
3) & October 2008: ``...break-up{[}.{]} BRUSSELS/AMSTERDAM - The government
of the Netherlands nationalized the Dutch banking and insurance activities
of troubled financial services company Fortis FOR.BRFOR.AS...''\footnotemark
\tabularnewline
4) & October 2008: ``'...euros' (\$970 million) worth of Dutch ABN AMRO
assets to Deutsche Bank last week. Fortis needed to sell some operations
to meet European Commission antitrust...''\footnotemark
\tabularnewline
5) & October 2008: ``bailout rescue for the Belgian-Dutch group on Sunday, a regulatory filing showed. ABN AMRO Netherlands profit rose by 1 million euros compared to the first''\footnotemark
\tabularnewline
6) & October 2009: ``...Fortis Bank Nederland. Under the deal, 15 months
and two ABN owners in the making, the acquisitive German lender will
boost its Dutch operations by...''\footnotemark
\tabularnewline
7) & December 2009: ``Julien Ponthus{[}.{]} BRUSSELS, Dec 1 (Reuters)
- BNP Paribas (BNPP.PA) on Tuesday raised its forecast for savings
from the integration of former Fortis assets''\footnotemark
\tabularnewline

\end{tabular}
\\
\protect\caption{Selected top-ranked excerpts for Fortis Bank 2007--2014}
\end{table}

\addtocounter{footnote}{-7} 
\stepcounter{footnote}\footnotetext{Ibid. /2008/09/25/idUSPRWP1420080925}
\stepcounter{footnote}\footnotetext{Ibid. /2008/09/29/markets-europe-stocks-open-idUSLT42646320080929}
\stepcounter{footnote}\footnotetext{Ibid. /2008/10/03/us-fortis-belgium-factbox-idUSTRE49289T20081003} 
\stepcounter{footnote}\footnotetext{Ibid. /2008/10/10/sppage012-la190235-oisbn-idUSLA19023520081010} 
\stepcounter{footnote}\footnotetext{Ibid. /2008/10/02/abnamro-idUSL220146920081002} 
\stepcounter{footnote}\footnotetext{Ibid. /2009/10/20/us-abnamro-idUSTRE59J1Y820091020} 
\stepcounter{footnote}\footnotetext{Ibid. /2009/12/01/bnp-fortis-idUSGEE5B00B420091201}

\section{Conclusions}

Starting from news text, on the one hand, and bank distress events,
on the other hand, we have presented a method for linking these two
types of data, in the form of a predictive model. The model provides
a coinciding stress index for banks over time and excerpts to describe
its drivers. Linking the unstructured and sparse text to the 243 distress
events is made possible by a deep learning setup that incorporates
distributed vector representation learning of word and document semantics.

We have demonstrated that our deep model is able to \emph{detect} coinciding
bank distress and provide a stress index for individual banks that can be studied at the aggregate European level as well. The model provides text excerpts to \emph{describe} the underlying events that are reflected in the index. As the current event data focus in particular on government interventions and state aid, the excerpts our model presents center around these topics to a large extent. A high index value indicates an increase in discussion related to distress and a particular bank. While neither the presence of such discussion nor elevated index values should be blindly interpreted as negative, the index and descriptions may serve to initiate and guide more thorough investigation.

As this paper has introduced the detect-and-describe approach, several interesting directions for future work open up. Variations to the compositional semantic modeling and heuristics for extraction of descriptions deserve further exploration. In particular, the final step of producing excerpts from word weights in a manner that supports interpretation well is all but trivial. Lastly, by extending the scope of the event data, we expect our approach to yield yet more interesting results.

\section*{Acknowledgment}

The authors are grateful to Filip Ginter, J\'ozsef Mezei and Tuomas Peltonen for their helpful comments. The paper also has benefited from presentation at the Systemic Risk Analytics and Macro-prudential Policy session of the Finnish Economic Association XXXVII Annual Meeting (KT-p\"aiv\"at), 12.2.2015, in Helsinki, Finland.



%


\bibliographystyle{plain}
\bibliography{references}

\begin{thebibliography}{10}

\bibitem{baroni2014don}
M.~Baroni, G.~Dinu, and G.~Kruszewski.
\newblock Don't count, predict! a systematic comparison of context-counting vs.
  context-predicting semantic vectors.
\newblock In {\em Proceedings of the 52nd Annual Meeting of the Association for
  Computational Linguistics}, volume~1, pages 238--247, 2014.

\bibitem{betz2014predicting}
F.~Betz, S.~Opric{\u{a}}, T.~A. Peltonen, and P.~Sarlin.
\newblock Predicting distress in european banks.
\newblock {\em Journal of Banking \& Finance}, 45:225--241, 2014.

\bibitem{cole_predicting_1998}
R.~A. Cole and J.~W. Gunther.
\newblock Predicting bank failures: A comparison of on- and off-site monitoring
  systems.
\newblock {\em Journal of Financial Services Research}, 13:103--117, 1998.

\bibitem{Groppetal2006}
R.~Gropp, J.~Vesala, and G.~Vulpes.
\newblock Equity and bond market signals as leading indicators of bank
  fragility.
\newblock {\em Journal of Money, Credit and Banking}, 38(2):399--428, 2006.

\bibitem{harris1954distributional}
Z.~S. Harris.
\newblock Distributional structure.
\newblock {\em Word}, 10(23):146--162, 1954.

\bibitem{le2014distributed}
Q.~Le and T.~Mikolov.
\newblock Distributed representations of sentences and documents.
\newblock In {\em Proceedings of the 31st International Conference on Machine
  Learning (ICML-14)}, pages 1188--1196, 2014.

\bibitem{lischinsky2011discourse}
A.~Lischinsky.
\newblock In times of crisis: a corpus approach to the construction of the
  global financial crisis in annual reports.
\newblock {\em Critical Discourse Studies}, 8(3):153--168, 2011.

\bibitem{mannasoo_explaining_2009}
K.~M\"{a}nnasoo and D.~G. Mayes.
\newblock Explaining bank distress in {E}astern {E}uropean transition
  economies.
\newblock {\em Journal of Banking \& Finance}, 33:244--253, 2009.

\bibitem{mikolov2013efficient}
T.~Mikolov, K.~Chen, G.~Corrado, and J.~Dean.
\newblock Efficient estimation of word representations in vector space.
\newblock In {\em Proceedings of Workshop at International Conference on
  Learning Representations}, 2013.

\bibitem{milne2014}
A.~Milne.
\newblock Distance to default and the financial crisis.
\newblock {\em Journal of Financial Stability}, 12:26--36, 2014.

\bibitem{mitchell2010composition}
J.~Mitchell and M.~Lapata.
\newblock Composition in distributional models of semantics.
\newblock {\em Cognitive science}, 34(8):1388--1429, 2010.

\bibitem{Gregoryetal2014}
R.~Nyman, D.~Gregory, K.~Kapadia, P.~Ormerod, D.~Tuckett, and R.~Smith.
\newblock News and narratives in financial systems: exploiting big data for
  systemic risk assessment.
\newblock BoE, mimeo, 2015.

\bibitem{Peltonenetal2015}
T.~Peltonen, A.~Piloui, and P.~Sarlin.
\newblock Network linkages to predict bank distress.
\newblock {ECB Working Paper, No. 1828}, 2015.

\bibitem{RonnqvistSarlin2015}
S.~R\"onnqvist and P.~Sarlin.
\newblock Bank networks from text: Interrelations, centrality and determinants.
\newblock {\em Quantitative Finance}, forthcoming, 2015.

\bibitem{rumelhart1986learning}
Hinton G.~E. Rumelhart D.~E. and R.~J. Williams.
\newblock Learning representations by back-propagating errors.
\newblock {\em Nature}, 323(6088):533--536, 1986.

\bibitem{Sarlin2013b}
P.~Sarlin.
\newblock On policymakers' loss functions and the evaluation of early warning
  systems.
\newblock {\em Economics Letters}, 119(1):1--7, 2013.

\bibitem{schmidhuber2015deep}
J.~Schmidhuber.
\newblock Deep learning in neural networks: An overview.
\newblock {\em Neural Networks}, 61:85--117, 2015.

\bibitem{Schutze:1992:DM:147877.148132}
H.~Sch\"{u}tze.
\newblock Dimensions of meaning.
\newblock In {\em Proceedings of the 1992 ACM/IEEE Conference on
  Supercomputing}, Supercomputing '92, pages 787--796, Los Alamitos, CA, USA,
  1992. IEEE Computer Society Press.

\bibitem{schutzePedersen1995irwordsenses}
H.~Sch\"utze and J.~Pedersen.
\newblock Information retrieval based on word senses.
\newblock In {\em Proceedings of the 4th Annual Symposium on Document Analysis
  and Information Retrieval}, pages 161--175, 1995.

\bibitem{manning2013deeplearning}
R.~Socher and C.~Manning.
\newblock Deep learning for natural language processing (without magic).
\newblock Keynote at the 2013 Conference of the North American Chapter of the
  Association for Computational Linguistics: Human Language Technologies
  (NAACL2013). http://nlp.stanford.edu/courses/NAACL2013/.

\bibitem{soo2013quantifying}
C.~K. Soo.
\newblock Quantifying animal spirits: news media and sentiment in the housing
  market.
\newblock {\em Ross School of Business Paper No. 1200}, 2013.

\bibitem{vanRijsbergen1979}
C.J. Van~Rijsbergen.
\newblock {\em Information Retrieval}.
\newblock Butterworth, 2nd ed., 1979.

\bibitem{wang2014copula}
W.~Y. Wang and Z.~Hua.
\newblock A semiparametric gaussian copula regression model for predicting
  financial risks from earnings calls.
\newblock In {\em Proceedings of the 52nd Annual Meeting of the Association for
  Computational Linguistics (ACL)}, 2014.

\end{thebibliography}

\end{document}